\documentstyle[12pt]{article}
\newcommand{\beq}{\begin{equation}}
\newcommand{\eeq}{\end{equation}}
\newcommand{\bea}{\begin{eqnarray}}
\newcommand{\ena}{\end{eqnarray}}

\begin{document}
\vspace{11. cm}
DFPD/92/TH/36
{\begin{center} {\Large
\bf The Electroweak Phase Transition \\
\vspace{0.3cm}
in a Nonminimal Supersymmetric Model} \\
\vspace{1.2cm}
{\large Massimo Pietroni} \\
\vspace{0.1cm}
{\em Dipartimento di Fisica ``Galileo Galilei", Universit\`{a} \\
     di Padova, Via F.Marzolo 8, 35131 Padova, Italia } \\
\end{center}
\vspace{1cm}
\begin{abstract}
We study the electroweak phase transition in a supersymmetric version of the
Standard Model, in which a gauge singlet superfield is added to the Higgs
sector. We show that the order of the transition is determined by the trilinear
soft supersymmetry breaking terms rather than by the $O ( m^{3} T )$ term in
the 1-loop,
$T\neq0$ corrections. This fact removes the Standard Model upper bound on the
Higgs mass, $ m_{H} < 55 GeV$, coming from the requirement that baryon
asymmetry is not washed out by anomalous electroweak processes.
We perform a numerical analysis of parameter space including
in the effective potential top-stop contribution to 1-loop radiative
corrections. We find that this model is compatible with the preservation of
baryon asymmetry for masses of the lightest scalar up to about 170 GeV.
\end{abstract}
\newpage
{\bf 1. Introduction.}
\vspace{.5 truecm}\\

Several years ago Sakharov [1] realized that the observed baryon asymmetry of
the Universe (BAU) could have been generated at some early stage of cosmic
evolution at which three conditions were fulfilled: B-violating interactions,
C and CP violation, and a departure from thermodynamic equilibrium.
All these conditions are fulfilled in grand unification theories (GUT),
in which the
baryon asymmetry is generated in the \mbox{out-of-equilibrium}, B-violating,
decay of some superheavy boson [2]. However this scenario presents
severe complications. First of all, as noted by Bocharev, Kuzmin and
Shaposhnikov[3], any nonzero fermion number (B+L) created at the GUT epoch
is almost completely erased [4] by the anomalous electroweak \mbox{
B+L-violating} processes [5], which are in equilibrium down to a temperature
$T \sim 10^{2}$ GeV. So, if no B-L is created at the GUT phase transition,
then no baryon excess survives down to the Fermi scale.
Moreover, if \mbox{B-L-violating} interactions  ($\delta L=2$ Majorana masses
for neutrinos, R-parity violating interactions in SUSY models [5]...)
are present and in equilibrium at high ($T\gg 100$ GeV) temperature, then also
an eventual B-L component of the GUT-generated BAU vanishes before the
onset of the electroweak era.

The above considerations led many people to investigate the possibility
that the BAU was generated at the electroweak phase transition,([6]-[13]).
Here the difficulty is threefold; first, CP violation in the Standard Model
is too small [7], second, it is not obvious that the
transition is of first order strongly enough to make the baryon number
production effective and, third, we must
avoid the wiping out of the baryon asymmetry by the anomalous processes.
The efficiency of
the latter is suppressed by the exponential of (minus) the `sphaleron' mass
[6] over the temperature
\beq \frac{E^{sph}(T)}{T} = \frac{4\pi v(T)}{g_{W}}\frac{B}{T}, \eeq
where $B(\frac{\lambda}{g_{W}^{2}})$ is a slowly varying function of the
couplings $B(0)=1.56$, $B(\infty)=2.72$, and $v(T)$ the value of the vacuum
expectation value (VEV) of the Higgs field at the temperature T.
The requirement that (1) is sufficiently large at the phase transition gives
a lower bound [6] on the value of the vacuum expectation value (VEV) at the
critical temperature (which
we define here as the temperature at which the effective potential becomes
flat at the origin, $V''(0)=0$)
\beq \frac{v(T_{C})}{T_{C}}\stackrel{>}{\sim} 1.3 . \eeq

In the Standard Model $v(T_{C})/T_{C}$ is given by the ratio of the
coefficient of the term cubic in the Higgs field, $\delta$, and the quartic
self-coupling, $\lambda$. So it is easy to see that (2) imposes an upper
bound on the Higgs mass
\beq \left( \frac{M_{H}}{246.2 GeV}\right)^{2}
\stackrel{<}{\sim}2.3 \: \delta . \eeq

The crucial $\delta$ coefficient is zero in the Standard Model at the tree
level; it is generated by the \mbox{one-loop} finite temperature corrections
[14,15] as a subleading term in the $M/T$ expansion [15], and its
evaluation requires a careful resummation of the leading infrared (IR)
divergencies
[14,16].
Despite all the uncertainties related to this procedure, it seems by now
established that the Standard Model bound (3) has been ruled out by LEP
results, $M_{H}>57$ GeV [17], or it will be in the near future.

Since $\delta$ receives contribution by the bosonic degrees of freedom with
masses lighter than the temperatures in consideration
\footnote {in fact it may be shown
numerically that the $M/T$ expansion is a good approximation up to
$\frac{M}{T}\sim 2.2$ [9]},
attempts have been made to relax the bound on $M_{H}$ by considering the
Standard Model
with an extended Higgs sector [6,8,9,18], or minimal SUSY Standard Model
(MSSM), in which two higgs
doublets are present [10]. In all these cases the cubic term is again a
\mbox{one-loop} effect and its evaluation meets the same IR problems that
we find in the Standard Model. Moreover, as discussed in ref. [13], there
is a large suppression
$(O (\alpha_{W}^{8}))$ of the baryon asymmetry in models containing only scalar
doublets, without extra CP violation sources than the Higgs potential and
the Yukawa couplings.

In this paper we study the electroweak phase transition in the simplest
possible
extension of the MSSM with soft
supersymmetry breaking, in which a gauge singlet superfield is added to the
superpotential [19,20]. From the point of view of the phase transition the main
feature of this model is the occurrence of trilinear soft SUSY breaking terms
in the Higgs sector of the tree level effective potential. These terms behave
as effective cubic terms in the radial direction from the origin of the field
space to the vacuum, so that the potential has a barrier along this direction
already at the tree-level. Moreover, including the leading (no IR problem)
finite temperature corrections, at \mbox{one-loop}, we get also positive
linear terms which enhance
the barrier between the false and the true vacuum.
As a result, this model leads very naturally to a strongly first order phase
transition and, as we will see, the bound (2) may be fulfilled in a
large portion of the parameter space corresponding to values of the mass of the
lighter scalar up to about 170 GeV \vspace{1. truecm}.\\
{\bf 2. The model.}
\vspace {.5 truecm}\\

We assume [20] that the squark and slepton fields have vanishing
vacuum expectation values so that we can restrict our attention to the Higgs
sector of the
superpotential involving the superfields $\tilde{H}_{1}$, $\tilde{H}_{2}$ and
$\tilde{N}$ only
 \[ W_{Higgs}=\lambda \tilde{H}_{1}\tilde{H}_{2}\tilde{N}-
\frac{1}{3}k\tilde{N}^{3}. \]

The tree level scalar potential is given by
\bea
V & = & V_{F}+V_{D}+V_{Soft} ;\\ \vspace{0.5 truecm}
V_{F} & = & |\lambda|^{2}\left[|N|^{2}(|H_{1}|^{2}+|H_{2}|^{2})
+|H_{1}H_{2}|^{2}\right]
       + k^{2} |N|^{4} -(\lambda k^{*} H_{1}H_{2}{N^{2}}^{*} + h.c);
\nonumber \\ \vspace{0.5 truecm}
V_{D} & = & \frac{g^{2}+g'^{2}}{8} (|H_{2}|^{2}-|H_{1}|^{2})^{2} +
\frac{g^{2}}{2}|H_{1}^{\dag}H_{2}|^{2} ;\nonumber \\ \vspace{0.5 truecm}
V_{Soft} & = & m_{H_{1}}^{2}|H_{1}|^{2} + m_{H_{2}}^{2}|H_{2}|^{2} + m_{N}^{2}
|N|^{2} \nonumber \\ &  & -(\lambda A_{\lambda}H_{1}H_{2}N + h.c.)
-(\frac {1}{3} k A_{k} N^{3} + h.c.) , \nonumber \ena

where $ H_{1}\equiv \left (\begin{array}{c}
                 H_{1}^{0}  \\
                 H^{-} \end{array}\right ), H_{2}\equiv \left (\begin{array}{c}
                                               H^{+} \\
                                               H_{2}^{0} \end{array} \right ).
$

Note in particular the presence of trilinear terms in $V_{Soft}$, which are
absent in the Higgs sector of the MSSM.

Redefining the global phases of $H_{1}$ and $N$ we can always take
$\lambda A_{\lambda}$ and $k A_{k}$ real and positive, while by an
$\mbox{SU(2)XU(1)}$ global rotation we put
\[
 v^{+} \equiv <H^{+}> =0   ,    \:
 v_{2} \equiv <H_{2}^{0}> \in R^{+} .
\]

We will not discuss the CP violation aspect of the BAU generation problem,
so we assume CP conservation. Choosing $\lambda k >0$ we avoid explicit CP
violation and find three degenerate minima (note that the problem has a
discrete $Z_{3}$ symmetry[21]) corresponding to different values for
$\phi_{1}$ and $\phi_{x}$, which are respectively the phases of
$v_{1}\equiv <H_{1}>$ and $x\equiv <N>$
\beq \left\{ \begin{array}{lccclcc}
        \phi_{1} & = & 0 &, & \phi_{x} & = & 0 ;\\
        \phi_{1} & = & \frac {2}{3}\pi &, & \phi_{x} & = & \frac{4}{3}\pi ;\\
        \phi_{1} & = & \frac {4}{3}\pi &, & \phi_{x} & = & \frac{2}{3}\pi .
\end{array}
\right. \eeq

Due to our assumption of CP conservation,
the true vacuum of the theory is the one with both phases equal to zero.

We will work in the unitary gauge
\[ H_{1}= \left ( \begin{array}{c}
                (Re\,H_{1}^{0}+\frac{i}{\sqrt{2}} \sin \beta\: A^{0})\\
                \sin \beta \: C^{+\,*}
\end{array} \right )\:,\:H_{2}=\left ( \begin{array}{c}
                                        \cos \beta \: C^{+}\\
                        (Re\,H_{2}^{0}+\frac{i}{\sqrt{2}} \cos \beta\: A^{0})
\end{array} \right ) ,
\]

where, as usual, $tg\beta =v_{2}/v_{1}$, and
\bea
 A^{0} & \equiv & \sqrt{2} (\sin\beta\: Im\, H_{1}^{0}+\cos\beta\:Im\,
H_{2}^{0}) ; \nonumber \\
C^{+} & \equiv & \cos \beta\: H^{+}+\sin \beta H^{-\,*}. \nonumber
\ena
The combinations orthogonal to $C^{+}$ and $A^{0}$ are the would-be Goldstone
bosons giving masses to the $W^{\pm}$ and $Z^{0}$. The physical degrees of
freedom of the model are given by three scalars, two pseudoscalars and one
charged scalar.

Before proceeding in the calculation of the finite temperature corrections,
we wish to comment on some aspects of the tree level potential which will be
used in what follows. First, it contains seven free parameters: $ \lambda$,
$ k$, $A_{\lambda}$, $A_{k}$, $m_{H_{1}}^{2}$, $m_{H_{2}}^{2}$ and
$m_{N}^{2}$.
Imposing the stationarity conditions in $(H_{1}^{0},H_{2}^{0},N)=
(v_{1},v_{2},x)$
with the constraint \[ v_{1}^{2}+v_{2}^{2}\equiv v^{2} =(246.2 \; GeV)^{2} , \]
we express the soft masses in terms of the six parameters $\lambda$, $k$,
$A_{\lambda}$, $A_{k}$, $tg\beta$ and $x$
\beq
   \left\{ \begin{array}{ccl}
         m_{H_{1}}^{2} & = & \lambda(A_{\lambda}+kx)x tg\beta
                       -\lambda^{2}(x^{2} + v^{2} \sin^{2}\beta) \\
                       & - & \frac{g^{2}+g'^{2}}{4} v^{2} \cos 2\beta
\vspace{0.5 truecm}\\
         m_{H_{2}}^{2} & = & \lambda (A_{\lambda}+kx)\,x\, cotg\beta
                       -\lambda^{2} (x^{2} + v^{2} \cos^{2}\beta) \\
                       & + & \frac{g^{2}+g'^{2}}{4} v^{2} \cos 2\beta
\vspace{0.5 truecm}\\
             m_{N}^{2} & = & \lambda A_{\lambda}\frac{v^{2}}{2x}\sin 2\beta
                        + k A_{k} x -\lambda^{2} v^{2} - 2 k^{2} x^{2} \\
                        & + & \lambda k v^{2} \sin 2\beta
    \end{array} \right.
\eeq
Note that (6) does not guarantee that $(v_{1},v_{2},x)$ is the global minimum
of the effective potential (4);
for each choice of the parameters we must verify that this is indeed the case.

In addition, we have to be sure that $v^{-}$ vanishes in the vacuum;
looking at the potential along the charged direction,
\bea V_{ch} & = & |H^{-}|^{2} \left[ m_{H_{1}}^{2} + \lambda^{2}|N|^{2} +
\frac{g^{2}}{4}
\left(|H_{1}|^{2}+|H_{2}|^{2}\right) \right. + \nonumber \\
 & & \left. \frac{g'^{2}}{4}(|H_{1}|^{2}-|H_{2}|^{2}) \right]
+\frac{g^{2}+g'^{2}}{8}|H^{-}|^{4} , \nonumber
\ena
we see that a sufficient condition in order to have $v^{-}=0$ as a global
minimum is
\beq m_{H_{1}}^{2}+\lambda ^2 x^{2}+\frac{g^{2}}{4}v^{2}+\frac{g'^{2}}{4}v^{2}
\cos 2\beta > 0
\eeq

Now we introduce the finite temperature corrections at the one-loop level.
As usual [15] they
can be split into $T=0$ and $T\neq 0$ contributions.

The former are given by ([22])
\beq
\Delta V^{1-loop}_{T=0} = \frac{1}{64 \pi^{2}} Str \left \{ M^{4}(\phi)
\left [ln \frac{M^{2}(\phi)}{Q^{2}} - \frac {3}{2} \right ] \right \} ,
\eeq
where $M^{4}(\phi)$, with $\phi=(H_{1},H_{2},N)$, is the field dependent mass
matrix and $Q$ the renormalization point. The $Q^{2}$ dependence in (8)
is compensated by that of the renormalized parameters, so that the full
effective potential is independent of $Q^{2}$.

We have approximated expression
(8) by considering only the top and stop contributions: in addition,
we take the two stops' masses to be degenerate
\beq
m^{2}_{\tilde{t}_{L,R}}\approx \tilde{m}^{2}_{Q}+h_{t}^{2}|H_{2}|^{2}\; \; ,\;
\;
m_{t}^{2}= h_{t}^{2}|H_{2}|^{2} ,
\eeq
where $\tilde{m}^{2}_{Q}$ is the squarks soft mass (we assume
$\tilde{m}^{2}_{U}=\tilde{m}^{2}_{Q}$) and $h_{t}$ is the Yukawa coupling.
We fix the renormalization scale $Q$ at a value $ \overline{Q} $ such that
the tree
level relations (6), expressed in terms of the renormalized parameters,
do not change, i.e.
\beq
\left.
\frac {\partial \Delta V_{T=0}^{1-loop}}{\partial H_{2}}
\right |_{\footnotesize{\begin{array}{c}
                Q^{2}={\overline{Q}}^{2} \\
                H_{2}=v_{2} \\
        \end{array}}}
=0.
\eeq
Since we have neglected the $\tilde{t}_{L}-\tilde{t}_{R}$ mixing terms in the
stop mass matrix, the correction (8) does not affect the pseudoscalar mass
matrix [22], while it modifies the scalar spectrum.

The $T\neq 0$ part of one-loop corrections [15] is given by the expression
\bea \Delta V^{1-loop}_{T\neq 0}(\phi,T) & = &
\sum_{F} \frac{g_{F}T^{4}}{2\pi^{2}} \int_{0}^{\infty} dx\:x^{2}
ln\left[ 1+e^{-\left( x^{2}+\frac{m_{F}^{2}(\phi)}{T^{2}}\right)^{\frac{1}{2}}}
\right] \\
 & + &
\sum_{B} \frac{g_{B}T^{4}}{2\pi^{2}} \int_{0}^{\infty} dx\:x^{2}
ln\left[ 1-e^{-\left( x^{2}+\frac{m_{B}^{2}(\phi)}{T^{2}}\right)^{\frac{1}{2}}}
\right]
\ena
where $m_{F(B)}^{2}(\phi)$ is the tree-level mass of a fermion (boson) in
presence of the background fields $\phi (=H_{1},H_{2},N)$, $g_{F(B)}$ is the
corresponding number of degrees of freedom and the sum runs over all
fermions (F) and bosons (B) of the theory.
For values of the fields such that $m(\phi)/T < 1$ we can expand
(11, 12) as ([15])
\beq
\begin{array}{ccl}
 \Delta V^{1-loop}_{T\neq 0}(\phi,T) & \approx &
 \sum_{F} g_{F} T^{4} \left[ -\frac{7\pi^{2}}{720}+\frac{m_{F}^{2}(\phi)}
{48 T^{2}} +O \left(\frac{m_{F}^{4}(\phi)}{T^{4}}ln\frac{m_{F}(\phi)}{T}
\right) \right] \\
 & + &  \sum_{B} g_{B} T^{4} \left[ - \frac{\pi^{2}}{90}+
\frac{m_{B}^{2}(\phi)}{24 T^{2}}
+O \left(\frac{m_{B}^{3}(\phi)}{T^{3}}\right) \right] ,
\end{array}
\eeq
where we have neglected the subleading $ m_{B}^{3}/T^{3}$
term in the expansion for the bosons. As we said, this term is essential in the
Standard Model as in the
MSSM in order to obtain a first order phase transition, while in the
present model
the barrier between the true and the false vacua is given mainly by the
tree-level trilinear terms.
We have checked numerically that the cubic term coming from one-loop
corrections
changes the values of $T_{C}$ and $V(T_{C})$ only by a few percent.

The degrees of freedom corresponding to masses $m>T$ are Boltzmann-suppressed
and in this limit the correction (15) reduces to ([9])
\[
\Delta V_{T\neq 0}^{1-loop}(\phi,T)\sim
\sum_{F,B} \frac{g_{i}T^{2}}{(2\pi)^{\frac{3}{2}}} m_{i}^{2}
\sqrt{\frac{T}{m_{i}}} e^{-\frac{m_{i}}{T}}
\left[ 1+\frac{15}{8}\frac{T}{m_{i}}+
O\left(\frac{T^{2}}{m_{i}^{2}}\right)
\right].
\]
Since we are considering temperatures of the order of $M_{W}$ the exponential
factor in the previous expression allows us, with good approximation,
to neglect SUSY particles
with masses above the SUSY threshold which we assume to be
\[
\tilde{m} \approx 1 TeV
\]
Lighter particles will contribute to (13) even if it may happen that
$M(\phi)/T \stackrel{>}{\sim}1$ for some of them, at large values of the field.
However (see the footnote n. 1) this does not change matters essentially
with respect to the situation in which we use (11,12). Moreover the use of (13)
allows us to perform an analytic study of the critical temperature and of the
phase transition which can help us to understand the main properties of the
model,
otherwise hidden by a numerical evaluation of the integrals in (11,12).

In the following we will derive the mass matrix for the relevant degrees of
freedom of the theory on the background of all the scalar fields. We will
then discuss bounds on the parameter space coming from the requirement of a
correct
symmetry breaking pattern (i.e. no VEV's for $A^{0}$, $Im N$ and for $C^{+}$),
and on the LEP limits on chargino masses. Finally we will scan the allowed
parameter space, find the values of the critical temperatures, and the
minimum of the potential at those temperatures.
\vspace{1. truecm}\\
{\bf 3. The mass matrix.}
\vspace{0.7 truecm}\\

Let us consider the matrix of the second derivatives of the tree-level
effective
potential: due to C and CP conservation it is  a block-diagonal
6x6 matrix containing the 3x3 neutral scalars matrix, the 2x2
pseudoscalar matrix and the charged scalar mass. The elements of these matrices
are given by:
\vspace{.7 truecm}\\
a)neutral scalars:
\vspace{.5 truecm}\\
in the basis $(Re H_{1},Re H_{2},Re N)$ we have
\\
\vspace{0.5 truecm}
\[
\begin{array}{ccl}
M_{S\:11}^{2} & = & 2 m_{H_{1}}^{2} +  2 \lambda^{2}(|N|^{2}+|H_{2}^{0}|^{2})\\
 & + & \frac{g^{2}+g'^{2}}{2} [Re({H_{1}^{0}}^{2})+ 2 |H_{1}^{0}|^{2}-
|H_{2}^{0}|^{2} - |H^{+}|^{2} + |H^{-}|^{2}] + g^{2} |H^{+}|^{2} ;
\end{array}
\]
\vspace{0.3 truecm}\\
\[
\begin{array}{ccl}
M_{S\:22}^{2} & = & 2 m_{H_{2}}^{2} +  2 \lambda^{2}(|N|^{2}+|H_{1}^{0}|^{2})\\
 & + & \frac{g^{2}+g'^{2}}{2} [Re({H_{2}^{0}}^{2})+ 2 |H_{2}^{0}|^{2}-
|H_{1}^{0}|^{2} - |H^{-}|^{2} + |H^{+}|^{2}] + g^{2} |H^{-}|^{2} ;
\end{array}
\]
\vspace{0.3 truecm}\\
\[
\begin{array}{ccl}
M_{S\:33}^{2} & = & 2 m_{N}^{2} +  2
\lambda^{2}(|H_{1}^{0}|^{2}+|H_{2}^{0}|^{2}
+|H^{+}|^{2}+|H^{-}|^{2}) + 4 k^{2}\left[ 3 (Re N)^{2}+(Im N)^{2}\right]\\
 & - &4 k A_{k} ReN -4 \lambda k Re (H_{1}H_{2}) ;
\end{array}
\]
\vspace{0.3 truecm}\\
\[
\begin{array}{ccl}
M_{S\:12}^{2} & = & [4 \lambda^{2} - 2 (g^{2}+g'^{2})] Re H_{1}^{0}
Re H_{2}^{0} -
4 \lambda A_{\lambda} Re N \\
 & - &4 \lambda k \left[ (ReN)^{2} - (ImN)^{2}\right]- 4\left( \lambda^{2} -
\frac{g^{2}}{2}\right) Re(H^{+} H^{-}) ;
\end{array}
\]
\vspace{0.3 truecm}\\
\[
M_{S\:13}  =  4 \lambda^{2} Re H_{1}^{0} Re N -4 \lambda k Re(H_{2}^{0} N^{*})
- 2 \lambda A_{\lambda} Re H_{2}^{0} ;
\]
\vspace{0.3 truecm}\\
\[
M_{S\:23}  =  4 \lambda^{2} Re H_{2}^{0} Re N -4 \lambda k Re(H_{1}^{0} N^{*})
- 2 \lambda A_{\lambda} Re H_{1}^{0} ;
\]
\vspace{0.5 truecm}\\
b)pseudoscalars:
\vspace{0.5 truecm}\\
in the basis ($A_{0}, ImN$)
\bea
M_{P\:11} & = & \sin^{2} \beta \left[ m_{H_{1}}^{2} + \lambda^{2} (|N|^{2} +
|H_{2}^{0}|^{2}) + \frac{g^{2}}{2} |H^{+}|^{2} \right. \nonumber \\
 & + & \left.  \frac{g^{2}+g'^{2}}{4} \left((2|H_{1}^{0}|^{2} - |H_{2}^{0}|^{2}
- Re ({H_{1}^{0}}^{2}) + |H^{-}|^{2} - |H^{+}|^{2}\right) \right] ;
\ena
\bea
M_{P\:22} & = &  m_{N}^{2} + +2 \lambda^{2} ( |H_{1}|^{2} + |H_{2}|^{2} )
\nonumber \\
 & + & 4 k^{2} [2 |N|^{2} - Re (N^{2})] +4 \lambda k Re(H_{1}H_{2}) +
4 k A_{k} Re N ;
\ena
\bea
M_{P\:12} & = & \frac{1}{\sqrt{2}} \sin \beta \left\{ 4 \lambda ^{2}
ImN ImH_{1}^{0} - 4 \lambda k Re (N^{*} H_{2}^{0})
+ 2 \lambda A_{\lambda} Re H_{2}^{0} \right\} \nonumber \\
 & + & \frac{1}{\sqrt{2}} \cos \beta \left\{ 4 \lambda ^{2}
ImN ImH_{2}^{0} - 4 \lambda k Re (N^{*} H_{1}^{0})
+ 2 \lambda A_{\lambda} Re H_{1}^{0} \right\} ; \nonumber
\ena
\vspace{0.5 truecm}\\
c)charged scalar:
\vspace{0.5 truecm}\\
\bea
m_{C}^{2} & = & \cos^{2} \beta \left[ m_{H_{2}}^{2} + \lambda^{2} (|N|^{2} +
|H^{-}|^{2}) +\frac{g^{2}-g'^{2}}{4} |H_{1}^{0}|^{2} \right. \nonumber \\
 & + & \left. \frac{g^{2}+g'^{2}}{4} (|H_{2}^{0}|^{2} + 2 |H^{+}|^{2} -
|H^{-}|^{2})
\right] \nonumber \\
& + & \sin \beta \cos \beta \left[ -\lambda^{2} (H_{1}^{0} H_{2}^{0} -
H^{+}H^{-})
 -\frac{g^{2}+g'^{2}}{4} (H^{+}H^{-})
\right.
\nonumber \\
 & + & \left. \frac{g^{2}}{2} (H_{1}^{0}H_{2}^{0}) + \lambda k N^{2} + \lambda
A
_{\lambda} N^{*} + h.c. \right] \nonumber \\
& + &  \sin^{2} \beta \left[ m_{H_{1}}^{2} + \lambda^{2} (|N|^{2} +
|H^{+}|^{2})+\frac{g^{2}-g'^{2}}{4} |H_{2}^{0}|^{2} \right. \nonumber \\
 & + & \left. \frac{g^{2}+g'^{2}}{4} (|H_{1}^{0}|^{2} + 2 |H^{-}|^{2} -
|H^{+}|^{2})
 \right] . \nonumber
\ena

Next we come to the SUSY particles. Among the squarks and leptons we should
consider only the stop-quarks $\tilde{t}_{L,R}$, the others having negligible
Yukawa  couplings. However $\tilde{t}_{L,R}$ masses (9) are dominated by
$\tilde{m}_{Q}$,
which at the electroweak scale is  much greater than $M_{W}$ (its
renormalization group equation is dominated by the
strong coupling [19]) and then their contribution is Boltzmann-suppressed.

Charginos have a 2x2 mass matrix (in the basis ($\tilde{H}^{-},
\tilde{W}^{-}$))
\beq
M_{ch}^{2}=
\left(
\begin{array}{cc}
\lambda N & g\sqrt{2}H_{2}^{0}\\
 g\sqrt{2}H_{1}^{0} & M_{2}
\end{array}
\right),
\eeq
where $M_{2}$ is the gaugino direct mass term.
Assuming
\[
M_{2} \approx \tilde{m} \gg M_{W}\approx \lambda x
\]
 we get
\[
{m_{ch1}}^{2} \approx \lambda ^{2} |N|^{2}\;\; ; \;\;
{m_{ch2}}^{2}\approx {M_{2}}^{2}+2 g^{2}(|H_{1}^{0}|^{2}+|H_{2}^{0}|^{2}) ,
\]
which shows that the lightest chargino may be in thermodynamic equilibrium
and contributes to (13). The LEP bound on the lowest
chargino mass [23]
\[
m_{\chi^{\pm}}>45 GeV
\]
will be imposed to constrain the parameter space for $\lambda$
and x.
The 5x5 neutralinos mass matrix, $M_{N}^{2}$
has two heavy (of order $M_{1,2}$) Boltzmann suppressed eigenvalues. With the
choice of the parameters that will be illustrated in the following,
the off-diagonal terms
${M_{N\;12}}^{2}$, ${M_{N\;13}}^{2}$ and ${M_{N\;23}}^{2}$, are negligible in
comparison with the three diagonal ones,
${M_{N\;11}}^{2}$, ${M_{N\;22}}^{2}$ and ${M_{N\;33}}^{2}$, so that the
approximate masses for
the three lightest neutralinos are given by the following expressions
\bea
        {m_{N1}}^{2} & \approx & \lambda ^{2}(|N|^{2}+|H_{2}^{0}|^2)
+(g^{2}+g'^{2})|H_{1}^{0}|^{2} ; \nonumber \\
        {m_{N2}}^{2} & \approx & \lambda ^{2}(|N|^{2}+|H_{1}^{0}|^2)
+(g^{2}+g'^{2})|H_{2}^{0}|^{2} ; \nonumber \\
        {m_{N3}}^{2} & \approx & \lambda ^{2}(|H_{1}^{0}|^{2}+|H_{2}^{0}|^2)
+4 k^{2}|N|^{2} . \nonumber
\ena

Now we are ready to write the one-loop, $T\neq 0$, correction in the
high temperature limit. It is given by
\bea
\Delta V^{1-loop}_{T\neq 0}(\phi,T) & = & \frac{T^{2}}{24}[Tr M_{S}^{2}+
Tr M_{P}^{2} + 2 m_{C}^{2} +6 M_{W}^{2} + 3 M_{Z}^{2} \nonumber \\
 & + &  6 m_{t}^{2} + 2 m_{ch1}^{2} + m_{N1}^{2}+ m_{N2}^{2}+ m_{N3}^{2}]' ,
\nonumber
\ena
where $[\cdots]'$ means that the sum runs on the light masses only. As the
field-dependent mass of a given particle becomes much greater than the
temperature,
the contribution of that particle to the previous expression is switched off.
Numerical evaluations [9] of the integrals in (11, 12) show that we can
approximate them at better than ten percent  using expression (13) for
masses up to
$m/T \sim 2$.
\newpage
{\bf 4. Determination of the critical temperature.}
\vspace{0.7 truecm}\\

We defined the critical temperature $T_{C}$ as that value of T at which
the origin of the field space becomes a saddle point for the effective
potential.\
In fact, it is well known [9,24] that the transition
proceeds by quantum tunnelling and is completed at a temperature higher than
$T_{C}$, however we are now interested in determining the order of the
transition, which is parameterized by the vacuum expectation
value  at $T_{C}$ ($v(T_{C})=0$ means a second order phase transition,
\mbox{$v(T_{C})\neq 0$} a first order one).

It is easy to check that at high temperature symmetry is restored, in the
sense that the origin is a minimum of the effective potential.\\
The critical temperature is then defined by the condition
\beq
    \left.
                det \left[{M_{S}^{T}}^{2}(T_{C})\right]
    \right |_{\phi_{i}=0}
                                =0 ,
\eeq
where the effective mass matrix $M^{T}_{S}$ is given by the second derivatives
of
the full one-loop potential at finite temperature with respect to the
scalar fields.
In the origin of the field space the effective mass matrix is approximatively
diagonal in the basis
$(Re H_{1},Re H_{2},Re{N})$
\beq
        \left. {M_{S}^{T}}^{2}(T_{C})\right|_{\phi_{i}=0} = 2
\left( \begin{array}{ccc}
m_{H_{1}}^{2} + \frac{T^{2}}{24}C_{H_{1}} & \frac{T^{2}}{24}C_{1 2} & 0 \\
\frac{T^{2}}{24}C_{1 2} & m_{H_{2}}^{2} + \frac{T^{2}}{24}C_{H_{2}} +
\Delta_{rad} & 0 \\
 0 & 0 & m_{N}^{2} + \frac{T^{2}}{24}C_{N}
\end{array} \right) ,
\eeq
where
\bea
C_{H_{1}} & = & \lambda^{2}(5+\cos^{2}\beta)+\frac{g^{2}}{4}(22-\cos 2\beta)
+\frac{g'^{2}}{4}(8-3 \cos 2\beta) ; \nonumber \\
C_{H_{2}} & = & 6h_{t}^{2}+ \lambda^{2}(5+\sin^{2}\beta)+\frac{g^{2}}{4}
(22+\cos 2\beta)+\frac{g'^{2}}{4}(8+3 \cos 2\beta) ; \nonumber \\
C_{1 2} & = & \sin2\beta(\frac{g^{2}}{2}-\lambda^{2}) ; \nonumber \\
C_{N} & = & 9 \lambda^{2}+12 k^{2} +3 \lambda k \sin2\beta , \nonumber
\ena
and
\[ \Delta_{rad}= \frac{3 h_{t}^{2}}{8 \pi^{2}} {\tilde{m}}^{2}
log\frac{ {\tilde{m}}^{2}}{{\tilde{m}}^{2}+m_{t}^{2}} .
\]

Neglecting the off-diagonal term, the critical temperature is given by the
highest among $T_{H_{1}}^{S}$, $T_{H_{2}}^{S}$ and $T_{N}^{S}$ where
\beq
   {T_{i}^{S}}^{2}=-\frac{24 m_{i}^{2} + \delta_{i, \: H_{2}} \Delta_{rad}}
{C_{i}} \;\;\;  (i=H_{1},H_{2},N).
\eeq
Obviously the effective potential can become flat only along those directions
corresponding to negative soft masses.\
Negative values of $m_{N}^{2}$ may however be dangerous for CP conservation,
which we have assumed as a `boundary condition' for the vacuum at T=0;
if we look at the effective mass matrix for pseudoscalars in the origin
of field space, and in the basis $(A^{0},ImN)$, we get
\[
      {M_{P}^{T}}^{2}=2
\left(
        \begin{array}{cc}
m_{H_{1}}^{2}\sin^{2}\beta+m_{H_{2}}^{2}\cos^{2}\beta+\frac{T^{2}}{24}P_{A^{0}}
& 0 \\
0 & m_{N}^{2}+\frac{T^{2}}{24}P_{N}
         \end{array}
\right)
\]
where
\bea
        P_{A^{0}} & = & \lambda^{2}(3 + \frac{5}{2} \sin^{2}2\beta) +
\frac{5}{16} (g^{2}+g'^{2})\cos^{2}2\beta \nonumber \\
        P_{N} & = & 9\lambda^{2} + 12 k^{2} -3\lambda k \sin2\beta. \nonumber
\ena
Then, at the temperature
\[
        {T_{N}^{P}}^{2}=-\frac {24 m_{N}^{2}}{P_{N}}\geq{T_{N}^{S}}^{2}
\]
the potential becomes flat, in the origin, along the direction $Im N$,
while it remains convex in the direction $Re N$.
So, if at that temperature the transition in the scalar fields has not yet
taken place
\footnote {Note [20] that in $(ReH_{1},ReH_{2},ReN)=(v_{1},v_{2}
,x)$ the point $(A^{0},ImN)=(0,0)$ is always a minimum, as the pseudoscalar
mass matrix has only positive eigenvalues.}
then the imaginary part of the fields acquires a vacuum expectation value, and,
when the temperature goes to zero, the vacuum is no more $\phi_{1}=\phi_{x}=0$
(see eq.(5)).

Due to the heavy top [19], $m_{H_{2}}^{2}$ can run to
negative values at low energy, whereas, if $\lambda$ is not too large,
$m_{H_{1}}^{2}$ remains positive. It seems also plausible that, for
small $\lambda$ and k also $m_{N}^{2}$ remains positive, although a
detailed resolution of the RGE's for the soft parameters would be necessary
in order to clarify this point.

However, in what follows, we will constrain the parameter space by means of
eq. (6), requiring that the only negative soft mass is $m_{H_{2}}^{2}$.

The critical temperature will then be given, in any case, by
\[
T_{C} \approx T^{S}_{H_{2}} = \sqrt{-\frac{24 m_{H_{2}}^{2} + \Delta_{rad}}
{C_{H_{2}}}} ,
\]
where the $\approx$ is due to the fact that we have neglected the off-diagonal
term in (16). Anyway, numerical results will be obtained by solving the
condition
(17) with the complete matrix.
\vspace{1. truecm}\\
{\bf 5. Minimization of the potential}
\vspace{0.7 truecm}\\

Once we have ensured the correct pattern of symmetry breaking, i.e. no VEV's
for $Im N, A^{0}$ and $C^{+}$ (this last condition is guaranteed by eq. (7))
we can study
the minimization of the potential, at the critical temperature, in the scalar
directions only.

Before illustrating the numerical results of this minimization let us give an
intuitive view of what actually happens, namely, of the crucial role played
by the trilinear soft SUSY breaking parameters $A_{\lambda}$ and $A_{k}$ in
the formation of a minimum of the effective potential for large values of
the fields ($\sim T_{C}$).\\
We define the polar coordinates
\[
\left\{
\begin{array}{lcl}
Re H_{1}^{0} & = & Y \cos\alpha \; \cos\beta^{T} ; \\
Re H_{2}^{0} & = & Y \cos\alpha \; \sin\beta^{T} ; \\
Re N         & = & Y \sin\alpha .
\end{array}
\right.
\]
The full effective potential may then be written as the sum of terms of
different order in ${\tilde{Y}}\equiv Y/T$
\beq
{\tilde{V}} \equiv \frac{V}{T^{4}}= a {\tilde{Y}} + b \tilde{Y}^{2}+
                c \tilde{Y}^{3} + d \tilde{Y}^{4}+\cdots ,
\eeq
where the ellipses indicate terms coming from
$\Delta V^{1-loop}_{T=0}$, which, in our approximation, are different from zero
in the direction $H_{2}^{0}$ only and, due to the renormalization condition
(10), will be small as long as we search a minimum at $T_{C}$
with $H_{2}^{0}$ of the same order of $v_{2}$.

The coefficients in (20) are given by:

\vspace{0.5 truecm}
i) linear terms:
\[
a = \frac{1}{8}\sin 2\beta \sin\alpha \frac{\lambda A_{\lambda}}{T} ;
\]

ii) quadratic terms:
\[
b=\frac{1}{T^{2}} \left[ \cos^{2}\alpha (m_{H_{1}}^{2} \cos^{2}\beta^{T}+
        m_{H_{2}}^{2} \sin^{2}\beta^{T}) + m_{N}^{2} \sin^{2}\alpha+
\frac{T^{2}}{24} (\cdots)\right] ;
\]
here we do not write explicitly the terms coming from finite temperature
corrections;
for the present purposes what matters is that the complete quadratic term
is positive or zero at $T=T_{C}$ in any direction;

\vspace{0.5 truecm}
iii) cubic terms:
\[
c=-\frac{1}{T} \sin\alpha \left( \lambda A_{\lambda}
\cos^2 \alpha \sin 2\beta^{T}
+ \frac{2}{3} k A_{k} \sin^{2} \alpha \right) ;
\]

iv) quartic terms:
\bea
d & = & \lambda^{2} \cos^{2} \alpha \left( \sin^{2} \alpha +
\frac{1}{4} \cos^{2}\alpha \sin^{2} 2\beta^{T} \right) + k^{2}\sin^{4} \alpha
\nonumber \\
& - & \frac{\lambda k}{4} \sin^{2} 2 \alpha \sin 2\beta^{T} +
\frac {g^{2}+g'^{2}}{8} \cos^{4} \alpha \cos^{2} 2\beta^{T} . \nonumber
\ena
At large values of the fields, ${\tilde{Y}}^{2}\geq a / c  (\sim O(1))$
(that is $Y\geq T_{C}$) the potential has a minimum at
\[ \tilde{Y}\approx -\frac{3 c}{4 d} ;
\]
the condition for this to be a global minimum is (neglecting the quadratic
term)
\beq
\left|
\frac {c^{3}}{a d^{2}}
\right|\sim \frac{ \sin^{2} \alpha \left( \lambda A_{\lambda}
\cos^2 \alpha \sin 2\beta^{T}
+ \frac{2}{3} k A_{k} \sin^{2} \alpha \right)^{3}}
{A_{\lambda} \sin 2\beta T_{C}^{2} \lambda^{5}}
\gg 1
\eeq
Assuming that eq.(21) is satisfied, the question is now to
determine the direction of the vacuum in the field space.
Since the only negative
term appearing in (20) is the cubic one, and it is proportional to $\sin
\alpha$
, there will be always a $Re N$-component in the direction of the vacuum.

However, for large
$A_{\lambda}$ the l.h.s. of (21) decreases for $\sin\alpha =1$ (it is the
effect of the linear term that raises the potential in that direction), so
that the $\sin \alpha \neq 0$ directions become favorite, and an electroweak
first-order (nonzero VEV's for $Re H_{1}, Re H_{2})$ phase transition takes
place.
\vspace{1. truecm}\\
{\bf 6. Numerical results.}
\vspace {0.5 truecm}\\

Contrary to the case of the Standard Model, the constraint (2) on the vacuum
expectation value of the fields at the critical temperature cannot
be written as a simple bound on the masses of the scalars in this model.
In fact, approximate analytic expressions for these masses can be obtained
[20] in the limiting cases $x\gg v_{1}, v_{2}$ or $x\ll v_{1}, v_{2}$,
but we are here interested in values of $x$ of the same order of the
electroweak VEV's $v_{1}, v_{2}$ \footnote {One of the original reasons
for the introduction of this model was the so called $\mu$-problem, whose
solution requires $\lambda x \sim O(M_{W})$ [20].}.
So we are induced to perform a numerical
investigation on the parameter space in order to find the region in which
the bound (2) is satisfied, and then look at the corresponding values for the
mass of the lightest scalar.\\
In addition to the six tree-level parameters, we have to consider also
those appearing in the one-loop corrections, namely, the
top Yukawa coupling $h_{t}$, the stop soft SUSY breaking mass
$\tilde{m}_{Q}^{2}$, and the gauginos direct masses $M_{1}$ and $M_{2}$ (14).
We will assume a common value for $\tilde{m}_{Q}^{2}$, $M_{1}^{2}$ and
$M_{2}^{2}$
\[
\tilde{m}_{Q}^{2}=M_{1}^{2}=M_{2}^{2}=\tilde{m}^{2} = 1 TeV .
\]
Then we constrain $h_{t}$, $\lambda$, and $k$ by means of the
renormalization group analysis performed by the authors of ref [25].
Requiring that the coupling $\lambda$ remains perturbative up to a large
(say $10^{16}$ GeV) scale, they find, at $M_{Z}$
\beq
\lambda^{2}(M_{Z})  <  \frac{2 M_{Z}^{2}}{v^{2}}  =  0.274 ,
\eeq\\
and
\vspace {.5 truecm}
\beq
h_{t}(M_{Z})       \approx  0.97
\eeq
Moreover, RGE's for the couplings have the fixed ratio point (neglecting
$g, g'$ with respect to $h_{t}$)
\beq
\frac{k^{2}}{\lambda^{2}}=\frac{1}{2}.
\eeq
We will fix $\lambda$ at its upper value as given by (22) and $h_{t}$ and
$k$ according to (23), (24).

For fixed $M_{2}$ and $\lambda$ the experimental limit [23] on the chargino
mass
gives a lower bound on x, approximatively
\beq
x \ge x_{min}\approx \frac{45 GeV}{\lambda}.
\eeq
We then fix $tan \beta$ to some typical value (we will use $tan \beta = 2,
10$) and require (see the discussion on the critical temperature)
\[
m_{H_{1}}^{2}>0, m_{H_{2}}^{2}<0.
\]
This, in turn, implies
\beq
A_{\lambda}^{min}<A_{\lambda}<A_{\lambda}^{max},
\eeq
where
\bea
A_{\lambda}^{min} & = & \frac{1}{x}\left[ \lambda cotg \beta \left(
x^{2}+\frac{v^{2}}{4} \right) + \frac {\lambda v^{2}}{4} \sin 2\beta -
k x^{2} \right] ; \nonumber \\
A_{\lambda}^{max} & = & \frac{1}{x}\left[ \lambda tg \beta \left(
x^{2}+\frac{v^{2}}{4} \right) + \frac {\lambda v^{2}}{4} \sin 2\beta -
k x^{2} \right] , \nonumber
\ena
or, equivalently
\[
\frac{\lambda}{\sin^{2} \beta} \left( x^{2}+ \frac{v^{2}}{4} \right)
< m_{c}^{2} < \frac{\lambda}{\cos^{2} \beta} \left( x^{2}+ \frac{v^{2}}{4}
\right) ,
\]
where $m_{c}^{2}$ is the mass of the charged scalar.
Eq. (26) automatically ensures that eq. (8) is satisfied, i.e., that the
charged fields do not acquire any VEV.

Finally, the requirement $m_{N}^{2} > 0$ implies a lower bound on $A_{k}$
\beq
A_{k} > 2 k x +\frac{\lambda v^{2}}{k x} \left[ \lambda - \sin 2\beta
\left( \frac{A_{\lambda}}{2 x} + k \right) \right] .
\eeq

For any triplet $(x, A_{\lambda}, A_{k})$ satisfying (25), (26), and (27),
we minimize numerically the effective potential
at zero temperature and verify that the vacuum is at $(Re H_{1}^{0},
Re H_{2}^{0}, Re N)=(v_{1}, v_{2}, x)$. We then calculate the critical
temperature $T_{C}$ according to the definition (17), and, finally,
minimize the effective potential at $T_{C}$.

In Fig. 1 we report $v(T_{c})/T_{c}$ against the lightest scalar ($m_{S_{1}}$)
mass for all the points that passed the above mentioned selections. As we see
there are a lot of points corresponding to $v(T_{c})/T_{c} > 1.3 $ and
$m_{S_{1}}>60 GeV$. Note that the maximum allowed values for $m_{S_{1}}$
reach $170 GeV$.

In Figs. 2a, 2b we plot the allowed region (i.e. the points corresponding to
$v(T_{c})/T_{c} > 1.3 $) in the
$m_{S_{1}}-m_{c}$ plane, where $m_{c}$ is the charged scalar mass;
$x$ is fixed at 300 GeV, while the upper and lower limits for $m_{c}$ are given
by (26).

Finally, in Fig. 3a, 3b we recover the result we got from the naive
discussion on the minimization of the potential, namely, the crucial role
played by $A_{\lambda}$ in the determination of the direction of the
transition. The dashed lines represent the values of $v(T_{C})$, while the
continuos ones those of $T_{C}$.

In conclusion, we have showed explicitly that minimal extensions of the MSSM
allow to easily circumvent the Higgs mass problem arising from the
study of the electroweak phase transition in the Standard Model and in the
MSSM. The main role is
played by the trilinear soft SUSY breaking terms, which are present in any
supersymmetric model with extra SU(2)-singlet superfields. Radiative effects
are also important as, due to the heavy top, they force the direction of the
transition (i.e. the flat direction of the potential at $T_{C}$)
to be always $Re H_{2}$.
\vspace{1. truecm}\\
{\bf Acknowledgments}
\vspace{0.7 truecm}\\
The author would like to thank A. Masiero, who inspired this work and followed
each stage of its preparation, and G.F. Giudice, who read the
manuscript and
provided helpful comments and suggestions. Useful discussions with D. Comelli,
F. Illuminati and F. Zwirner are also acknowledged.

\newpage
\begin{center}
\large{\bf
{References.}}\\
\end{center}
\vspace{1.5 truecm}
{\bf [1]} A.D. Sakharov, Pis'ma Zh. Eksp. Teor. Fiz. {\bf 5} (1967) 32,
JETP Lett. {\bf 5} 24 (1967)
\vspace{0.5 truecm}\\
{\bf [2]} A.Yu. Ignatiev, N.V. Krashnov, V.A. Kuzmin and A.N. Tavkhelidze,\\

Proc. Int. Conf. Neutrino '77 Vol.2 (Nauka, Moskow,1978)
p.293,\\

 Phys. Lett. {\bf B76 } (1978) 436;
\vspace{0.2 truecm}\\

M. Yashimura, Phys. Rev. Lett. {\bf 41}(1978) 281, {\bf 42} (1979) 476 (E);
\vspace{0.2 truecm}\\

S. Weinberg, Phys. Rev. Lett. {\bf 42}(1979) 850;
\vspace{0.2 truecm}\\

A.Yu. Ignatiev, V.A. Kuzmin and M.E. Shaposhnikov, Phys. Lett. {\bf B87} (1979)
 114 \vspace{0.5 truecm}\\
{\bf [3]} V.Kuzmin, V.Rubakov and M. E. Shaposnikov, Phys. Lett. {\bf B155} (
1985) 36
\vspace {0.5 truecm}\\
{\bf [4]} J.S. Harvey and M.S. Turner, Phys. Rev. {\bf D42} (1990) 3344
\vspace {0.5 truecm}\\
{\bf [5]} B. Campbell, S. Davidson, J. Ellis and K.A. Olive, Phis. Lett.
{\bf B256} (1991) 457;
 \vspace {0.2 truecm}\\

W.Fishler, G.F. Giudice, R.G. Leigh and S.Paban, Phys. Lett {\bf B258} (1991)
45;
\vspace {0.2 truecm}\\

K. Enqvist, A. Masiero and A. Riotto, Nucl. Phys. {\bf B373} (1992) 95
\vspace {0.5 truecm}\\
{\bf [6]} A.D. Dolgov, YITP/K 90 preprint (1991) submitted to Phys. Rep.;
\vspace{0.2 truecm}\\

A.I. Bochkarev and M.E. Shaposhnikov, Mod. Phys. Lett. {\bf A6}(1987) 417;
\vspace{0.2 truecm}\\

A.I. Bochkarev, S. Kuzmin and M.E. Shaposhnikov, Phys. Lett. {\bf B244} (1990)
275;
\vspace{0.2 truecm}\\

A.I. Bochkarev, S. Kuzmin and M.E. Shaposhnikov, Phys. Rev. {\bf D43} (1991)
369;
\vspace{0.5 truecm}\\
{\bf [7]} M.E. Shaposhnikov, Nucl.Phys {\bf B287} (1987) 757
\vspace {0.5 truecm}\\
{\bf [8]} N.Turok and J.Zadrozny, Nucl. Phys {\bf B369} (1992) 729
\vspace {0.5 truecm}\\
{\bf [9]} G.W. Anderson and L.J. Hall, Phys. Rev. {\bf D45} (1992) 2685
\vspace {0.5 truecm}\\
{\bf [10]} G.F. Giudice, preprint UTTG-35-91 \vspace{0.5 truecm} \\
{\bf [11]} M.Dyne, R.G. Leigh, P.Huet, A. Linde and D. Linde SLAC-PUB-5740,
\mbox{SLAC-PUB-5741}
\vspace{0.5 truecm}\\
{\bf [12]} A.Masiero and A.Riotto, University of Padova preprint DFPD /92/TH/22
\vspace{0.5 truecm}\\
{\bf [13]} M.Dine, P.Huet and R.Singleton Jr. Nucl. Phys {\bf B375} (1992) 625
\vspace{0.5 truecm}\\
{\bf [14]} J. I. Kapusta, Finite temperature field theory (Cambridge University
Press, 1989)
\vspace{0.5 truecm}\\
{\bf [15]} L.Dolan and R.Jackiw, Phys. Rev. {\bf D9} (1974) 3320;
\vspace{0.2 truecm}\\

S.Weinberg, Phys. Rev. {\bf D9} (1974) 3357
\vspace{0.5 truecm}\\
{\bf [16]} D.A. Kirznhits and A.D. Linde, Ann. Phys. {\bf 101} (1976) 195;
\vspace{0.2 truecm}\\

A.D. Linde, Rep. Progr. Phys. {\bf 42} (1979) 389;
\vspace{0.2 truecm}\\

K.Takahashi, Z.Phys. {\bf C26} (1985) 601;
\vspace{0.2 truecm}\\

D.E. Brahm, S.D.H. Hsu, preprint CALT-68-1705;
\vspace{0.2 truecm}\\

M.E. Carrington, preprint TPI-MINN-91/48-T;
\vspace{0.2 truecm}\\

M.E. Shaposhnikov, CERN-TH 6319/91
\vspace{0.5 truecm}\\
{\bf [17]} M. Davier in Proc. of the Lepton-Photon Symposium and Europhysics
Conference\\

on High Energy Physics, S. Hagerty, K. Petter and E. Quercigh (eds.)
(Geneva, 1991)
\vspace{0.5 truecm}\\
{\bf [18]} K. Enqvist, K. Kainulainen and I.Vilja,preprint NORDITA 92/10 P
\vspace{0.5 truecm}\\
{\bf [19]} J.P. Derendinger and C.A. Savoy, Nucl. Phys. {\bf B237} (1984) 307
\vspace{0.5 truecm}\\
{\bf[20]} J.Ellis, J.F. Gunion, H.E. Haber, L.Roszkowsky and F.Zwirner, \\

Phys. Rev. {\bf D39} (1983) 844
\vspace{0.5 truecm}\\
{\bf [21]} J.Ellis, K.Enqvist, D.V. Nanopoulos, K. Olive, M. Quiros and
F. Zwirner,\\

Phys. Lett. {\bf B176} (1986) 403
\vspace{0.5 truecm}\\
{\bf [22]} J. Ellis, G. Ridolfi, F. Zwirner, Phys. Lett. {\bf B262} (1991) 477
\vspace{0.5 truecm}\\
{\bf [23]} ALEPH collaboration, preprint CERN-PPE/91-149
\vspace{0.5 truecm}\\
{\bf [24]} A.D. Linde, Nucl. Phys. {\bf B216} (1983) 421, Phys. Lett. {\bf
B100}
 (1981) 137
\vspace{.5 truecm}\\
{\bf [25]} P.Bin\'eutruy, C.A. Savoy, Phys. Lett. {\bf B277} (1992) 453
\vspace{.5 truecm}\\

\newpage
\begin{center}
\large{\bf
{Figure Captions.}}\\
\end{center}
\vspace{1.5 truecm}
{\bf Fig. 1} $v(T_{C})/T_{C}$ ratio vs. the lightest scalar mass, for all the
allowed region for the parameters $\lambda$, k, x, $A_{\lambda}$, and $A_{k}$
(see text), and $tg\beta=2$.
\vspace{1.5 truecm}\\
{\bf Fig. 2a} Upper and lower bounds for the lightest scalar mass coming from
the various conditions imposed on parameter space (see text) plus the
requirement that $v(T_{C})/T_{C})>1/3$. x has been fixed to 300 GeV and
$tg\beta=2$.
\\
{\bf Fig. 2b} The same as in fig 2a with $tg\beta=10$.
\vspace{1.5 truecm}\\
{\bf Fig. 3a} $v(T_{C})$ and $T_{C}$ dependence from $A_{\lambda}$, for
$x=300 GeV$, $A_{k}=190 GeV$ and $tg\beta=2$.\\
{\bf Fig. 3b} Same as in Fig. 3a for
$x=400 GeV$, $A_{k}=300 GeV$ and $tg\beta=10$.

\end{document}